\documentclass[aps,prl,twocolumn,superscriptaddress,showpacs]{revtex4-1}
\usepackage{graphicx,amsmath,amssymb,bm}

\newcommand{\bra}[1]{\left\langle\, #1\,\right|}
\newcommand{\ket}[1]{\left|\, #1\,\right\rangle}
\newcommand{\kev}{\,\textrm{keV}}
\newcommand{\mev}{\,\textrm{MeV}}
\newcommand{\gev}{\,\textrm{GeV}}
\newcommand{\gevi}{\, \text{GeV}^{-1}}

\newcommand{\fmi}{\, \text{fm}^{-1}}
\newcommand{\fmiq}{\, \text{fm}^{-3}}
\newcommand{\kf}{k_{\rm F}}

\begin{document}

\title{Neutron matter at next-to-next-to-next-to-leading order in
chiral effective field theory}

\author{I.\ Tews}
\affiliation{Institut f\"ur Kernphysik,
Technische Universit\"at Darmstadt, 64289 Darmstadt, Germany}
\affiliation{ExtreMe Matter Institute EMMI,
GSI Helmholtzzentrum f\"ur Schwerionenforschung GmbH, 64291 Darmstadt, Germany}
\author{T.\ Kr\"uger}
\affiliation{Institut f\"ur Kernphysik,
Technische Universit\"at Darmstadt, 64289 Darmstadt, Germany}
\affiliation{ExtreMe Matter Institute EMMI,
GSI Helmholtzzentrum f\"ur Schwerionenforschung GmbH, 64291 Darmstadt, Germany}
\author{K.\ Hebeler}
\affiliation{Department of Physics, The Ohio State University, 
Columbus, OH 43210, USA}
\author{A.\ Schwenk}
\affiliation{ExtreMe Matter Institute EMMI,
GSI Helmholtzzentrum f\"ur Schwerionenforschung GmbH, 64291 Darmstadt, Germany}
\affiliation{Institut f\"ur Kernphysik,
Technische Universit\"at Darmstadt, 64289 Darmstadt, Germany}

\begin{abstract}
Neutron matter presents a unique system for chiral effective field
theory (EFT), because all many-body forces among neutrons are
predicted to next-to-next-to-next-to-leading order (N$^3$LO). We
present the first complete N$^3$LO calculation of the neutron matter
energy. This includes the subleading three-nucleon (3N) forces for
the first time and all leading four-nucleon (4N) forces. We find
relatively large contributions from N$^3$LO 3N
forces. Our results provide constraints for neutron-rich matter in
astrophysics with controlled theoretical uncertainties.
\end{abstract}


\maketitle

The physics of neutron matter ranges from universal properties at low
densities to the structure of extreme neutron-rich nuclei and the
densest matter we know to exist in neutron stars. For these extreme
conditions, controlled calculations with theoretical error estimates
are essential. Chiral EFT provides such a systematic expansion for
nuclear forces~\cite{RMP}. This is particularly exciting for neutron
matter and neutron-rich systems, because all three- and four-neutron
forces are predicted to N$^3$LO~\cite{nm}.

Neutron matter based on chiral EFT has been studied using lattice
simulations~\cite{NLOlattice} at low densities, $n \lesssim n_0/10$
(with saturation density $n_0=0.16 \fmiq$), and following an in-medium
chiral perturbation theory approach~\cite{Wolfram1,Wolfram2}, where low-energy
couplings are adjusted to empirical nuclear matter properties. In
addition, the renormalization group (RG) has been used to evolve
chiral EFT interactions to low momenta~\cite{PPNP}, which has enabled
perturbative calculations for nucleonic
matter~\cite{nm,nucmatt2}. While these constrain the properties of
neutron-rich matter to a much higher degree than is reflected in
neutron star modeling~\cite{nstar}, the dominant uncertainties are due
to 3N forces, which were included only to N$^2$LO. A consistent
inclusion of higher-order many-body forces is therefore key.

Here we present the first calculations at nuclear densities based
directly on chiral EFT interactions without RG evolution. To this end,
we have studied the perturbative convergence of chiral two-nucleon
(NN) potentials for neutron matter in detail, and found that the
available N$^2$LO and N$^3$LO potentials with lower cutoffs
$\Lambda=450-500 \mev$ are perturbative. This is supported by small
Weinberg eigenvalues at low energies indicating the perturbative
convergence in the particle-particle channel~\cite{PPNP}. In neutron
matter, it comes as a result of effective range effects~\cite{dEFT},
which weaken NN interactions at higher momenta, combined with weaker
tensor forces among neutrons, and with limited phase space at finite
density due to Pauli blocking~\cite{nucmatt1}.

\begin{figure}[b]
\begin{center}
\vspace*{-2mm}
\includegraphics[width=0.85\columnwidth,clip=]{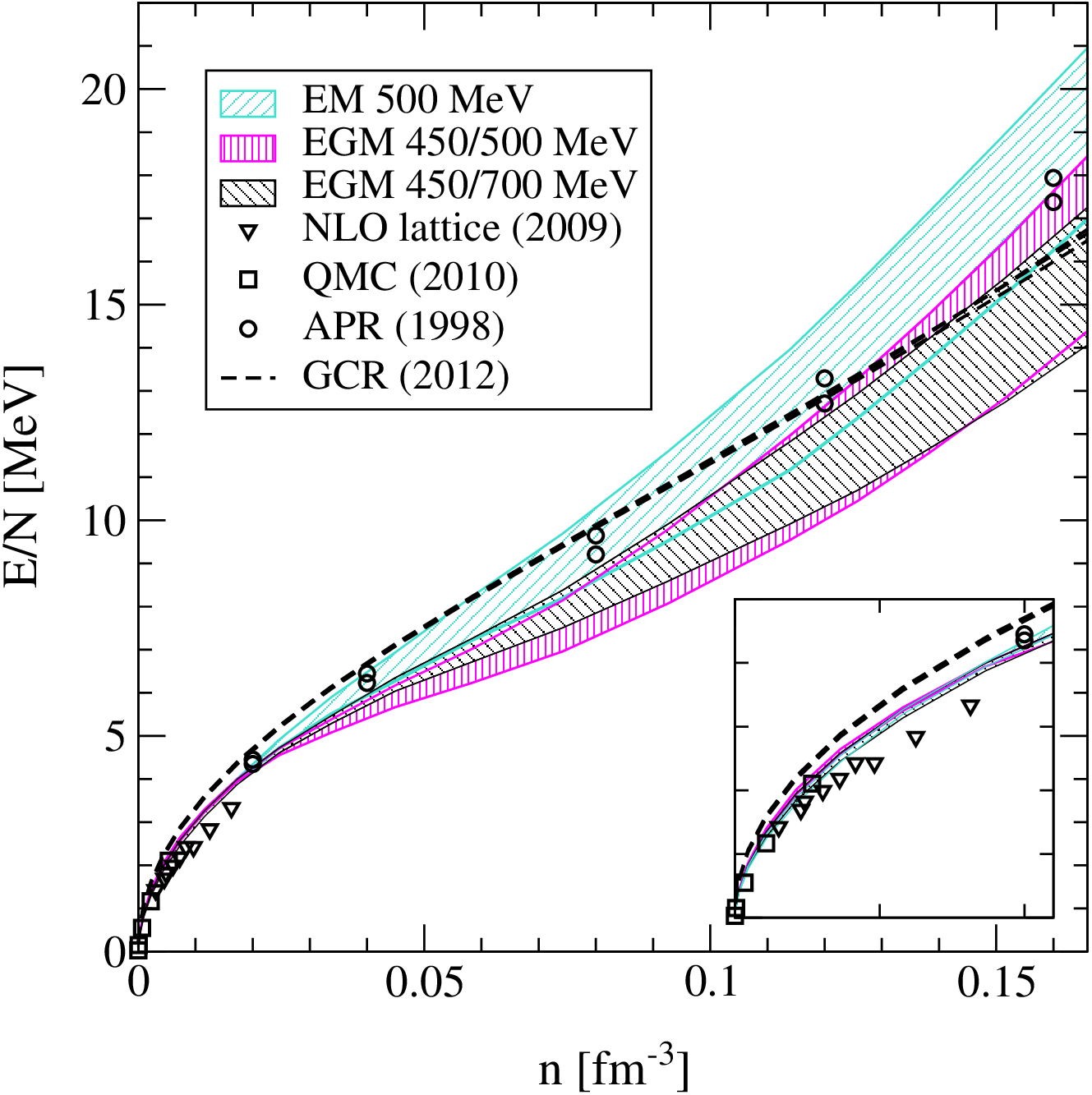}
\vspace*{-4mm}
\end{center}
\caption{(Color online) Neutron matter energy per particle as a
function of density including NN, 3N and 4N forces at N$^3$LO. The
three overlapping bands are labeled by the different NN potentials
and include uncertainty estimates due to the many-body
calculation, the low-energy $c_i$ constants and by varying the 3N/4N
cutoffs (see text for details).
For comparison, results are shown at low densities (see also the
inset) from NLO lattice~\cite{NLOlattice} and Quantum Monte Carlo
(QMC) simulations~\cite{GC}, and at nuclear densities from
variational (APR; the different points are with/without boost
corrections)~\cite{APR} and Auxiliary Field Diffusion MC
calculations (GCR)~\cite{GCR} based on adjusted nuclear force
models.\label{fig:N3LO}}
\end{figure}

At the NN level we use the N$^2$LO and N$^3$LO potentials developed by
Epelbaum, Gl\"ockle and Mei{\ss}ner (EGM)~\cite{EGM} with
$\Lambda/\widetilde{\Lambda} = 450/500$ and $450/700 \mev$
($\Lambda/\widetilde{\Lambda}$ denotes the cutoff in the
Lippmann-Schwinger equation and in the two-pion-exchange
spectral-function regularization, respectively).  We also use the
$\Lambda=500 \mev$ N$^3$LO NN potential of Entem and Machleidt
(EM)~\cite{EM}, which is most commonly used in nuclear structure
calculations. The larger $\Lambda=550 - 600 \mev$ NN potentials of
EGM and EM have been found to be nonperturbative~\cite{Note1} and are
therefore not included. Moreover, the LO NN contact couplings in the
$600/600$ and $600/700$ EGM potentials break Wigner symmetry
perturbatively (at the interaction level), with a repulsive
spin-independent $C_S$ and an unnaturally large spin-dependent $C_T
\sim C_S$, leading to unexpectedly large $C_T$-dependent 3N forces.

In this Letter, we include for the first time all N$^3$LO 3N and 4N
forces, which have been derived only
recently~\cite{IR,N3LOlong,N3LOshort,4N}, in addition to the N$^2$LO
3N forces. Figure~\ref{fig:N3LO} shows our complete N$^3$LO
calculation of the neutron matter energy as our main result, where the
bands include estimates of the theoretical uncertainties due to the
many-body calculation and in the many-body forces.

For neutrons, only the two-pion-exchange 3N forces contribute at
N$^2$LO~\cite{nm}. For the corresponding low-energy constants $c_1$
and $c_3$, we take the range of values from a high-order
analysis~\cite{Krebs2012}, at N$^2$LO: $c_1=-(0.37-0.81) \gev^{-1}$
and $c_3=-(2.71-3.40) \gev^{-1}$ (which includes the $c_i$ values in
the EGM and EM NN potentials), and when the N$^2$LO 3N forces are
included in an N$^3$LO calculation: $c_1=-(0.75-1.13) \gev^{-1}$ and
$c_3=-(4.77-5.51) \gev^{-1}$. It has been shown~\cite{nm} that the
N$^2$LO 3N force contributions in neutron matter can be to a good
approximation calculated at the Hartree-Fock level. 
In this first calculation, we therefore evaluate the N$^3$LO
3N and 4N force contributions to the energy per particle $E/N$ at the
Hartree-Fock level. The $A$-body contributions are then given by
\begin{multline}
\frac{E}{N} = \frac{1}{n} \frac{1}{A!} \sum\limits_{\sigma_1,\ldots,\sigma_A}
\int \frac{d{\bf k}_1}{(2\pi)^3} \, \cdots \int \frac{d{\bf k}_A}{(2\pi)^3}
\, f^2_R \, n_{{\bf k}_1} \cdots \, n_{{\bf k}_A} \\
\times \bra{1\ldots A} \mathcal{A}_{A} \sum\limits_{i_1 \neq \ldots \neq i_A=1}^A
V_{A}(i_1,\ldots,i_A) \ket{1\ldots A} ,
\end{multline}
with short-hand notation $i \equiv {\bf k}_i\sigma_i$. $\mathcal{A}_A$
denotes the $A$-body antisymmetrizer and $n_{{\bf k}_i}=\theta(k_{\rm
F} -k_i)$ the Fermi-Dirac distributions at zero temperature. We use a
Jacobi-momenta regulator; in terms of ${\bf k}_i$ given by $f_R =
\exp[-((k_1^2+\ldots+k_A^2-{\mathbf{k}_1\cdot\mathbf{k}_2}
-\ldots-\mathbf{k}_{A-1}\cdot\mathbf{k}_A)/(A \Lambda^2))^{n_\textrm{exp}}]$
with $n_\textrm{exp} =4$ and 3N/4N cutoff $\Lambda=2-2.5 \fmi$. For
the nucleon and pion mass, we use $m=938.92\mev$ and
$m_{\pi}=138.04 \mev$, and for the axial coupling
$g_A=1.29$ and the pion decay constant $f_{\pi}=92.4\mev$.

\begin{figure*}[t]
\begin{center}
\includegraphics[height=5.2cm,clip=]{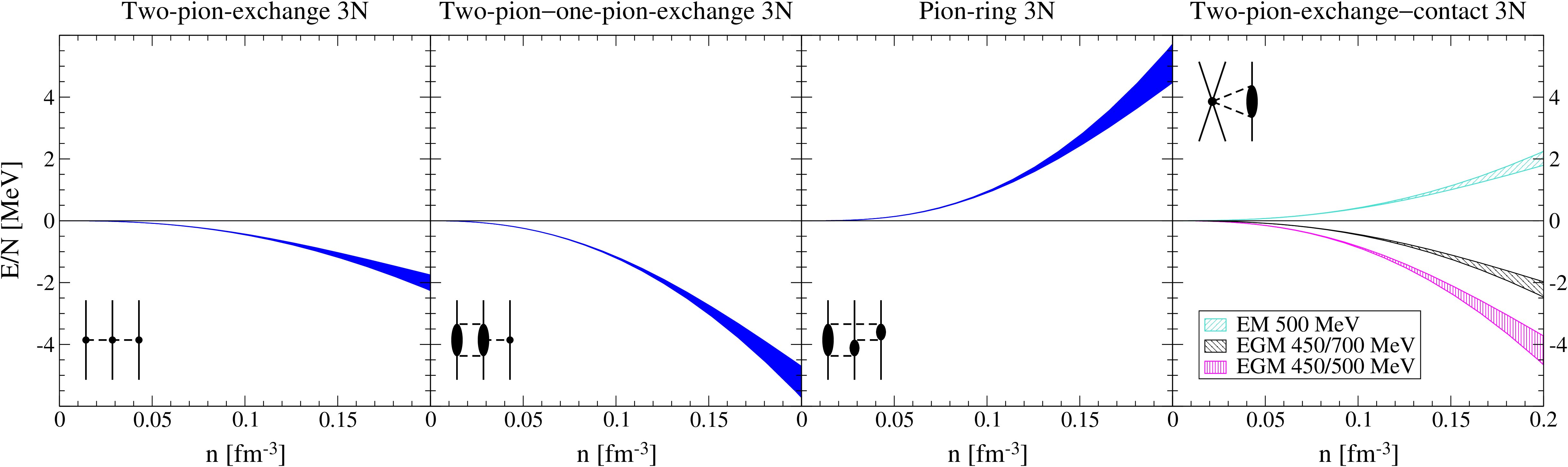}
\newline\newline
\vspace*{-2mm}
\includegraphics[height=5.2cm,clip=]{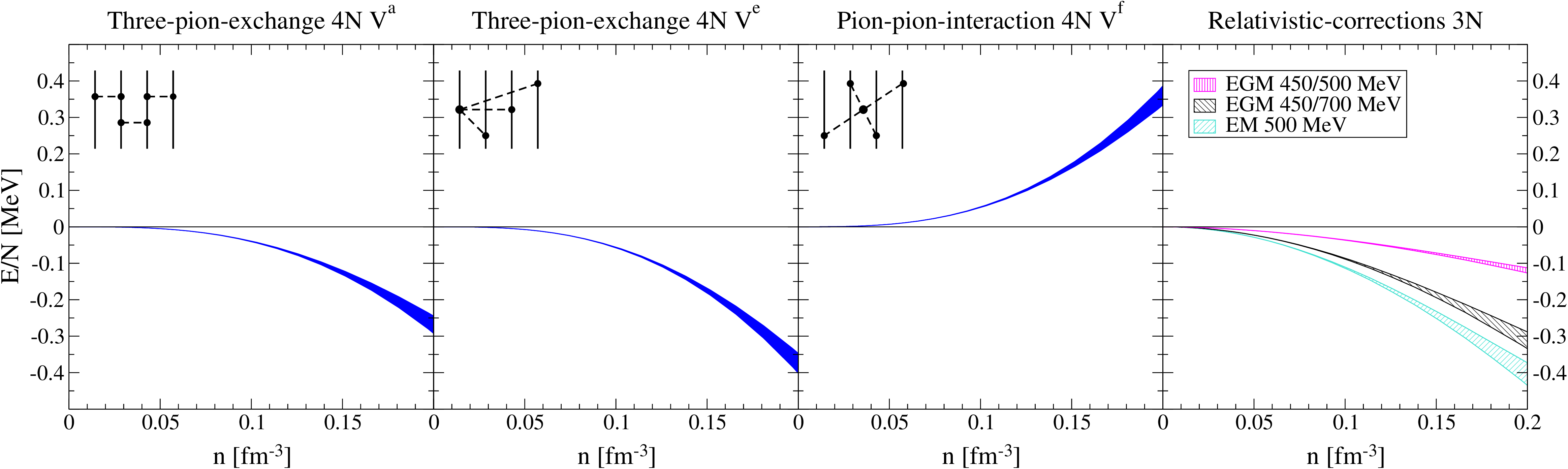}
\vspace*{-4mm}
\end{center}
\caption{(Color online) Energy per particle versus density for all
individual N$^3$LO 3N and 4N force contributions to neutron matter at the
Hartree-Fock level. The bands are obtained by varying the 3N/4N cutoff
$\Lambda = 2-2.5 \fmi$. For the two-pion-exchange--contact and the
relativistic-corrections 3N forces, the different bands correspond to
the different NN contacts, $C_T$ and $C_S$, determined consistently
for the N$^3$LO EM/EGM potentials. The inset diagram 
illustrates the 3N/4N force topology.\label{fig:individual}}
\vspace*{-2mm}
\end{figure*}

Chiral 3N forces at N$^3$LO can be grouped into
\begin{equation}
V^{\textrm{N$^3$LO}}_{\rm 3N} = V^{2\pi}+V^{2\pi\textrm{-}1\pi}+V^{\textrm{ring}}
+V^{2\pi\textrm{-cont}}+V^{1/m} \,,
\end{equation}
where we take the long-range parts, the subleading two-pion-exchange,
the two-pion--one-pion-exchange and the pion-ring 3N forces, from
Ref.~\cite{N3LOlong}, and the short-range parts, the
two-pion-exchange--contact and relativistic $1/m$-corrections 3N
forces from Ref.~\cite{N3LOshort}. In Fig.~\ref{fig:individual}, we
give the individual Hartree-Fock contributions to the neutron matter
energy. The evaluation is aided because parts of the different 3N
force topologies vanish for neutrons, and the results have been
checked by two independent calculations. The details of the
calculation will be presented in a future paper. At the Hartree-Fock
level, the 3N/4N contributions change by $< 5\%$ if the cutoff is
taken to infinity (i.e., $f_R=1$), but we will also include
N$^2$LO 3N forces beyond Hartree-Fock. This requires a consistently
used regulator. Estimates of the theoretical uncertainty are provided
by varying the 3N/4N cutoff.

The two-pion-exchange 3N forces at N$^3$LO can be largely written as
shifts of the low-energy constants, $\delta c_1 = -0.13 \gevi$ and
$\delta c_3 = 0.89 \gevi$~\cite{N3LOlong} of the N$^2$LO 3N forces,
plus a smaller contribution. The resulting energy of about $-1.5 \mev$
per particle at saturation density $n_0$ in Fig.~\ref{fig:individual}
is $\sim 1/3$ of the N$^2$LO 3N energy, as expected based on the
chiral EFT power counting.
In contrast, the two-pion--one-pion-exchange 3N force contributions,
which include 14 diagrams, are relatively large with $-3.6 \mev$ per
particle at saturation density. Of similar, but opposite size are the
pion-ring 3N force contributions, with $+3.3 \mev$ per particle at
$n_0$. The shorter-range parts of N$^3$LO 3N forces depend on the
momentum-independent NN contacts, $C_T$ and $C_S$, which we take
consistently from the N$^3$LO EM/EGM potential used. The contributions
from the two-pion-exchange--contact 3N forces include 11 diagrams and
depend only on $C_T$. The resulting energy ranges from $-2.8$ to $+1.3
\mev$ at $n_0$ depending on the NN potential used.  These larger 3N
results at N$^3$LO are consistent with contributions from the large
$c_i$ constants at N$^4$LO exactly in these three
topologies~\cite{Krebs2012}. This shows that higher-order many-body
forces still need to be investigated and that a chiral EFT with
explicit $\Delta$ excitations may be more efficient, since this would
capture these effects already at N$^3$LO. Finally, the
relativistic-corrections 3N forces depend also on $\bar{\beta}_8$ and
$\bar{\beta}_9$~\cite{N3LOshort} and contribute at the few hundred keV
level.

The 4N force contributions in Fig.~\ref{fig:individual} are an order
of magnitude smaller than those from the N$^3$LO 3N forces and of
similar size as the 3N relativistic corrections. We follow the 4N
force notation $V^a$ through $V^n$ of Ref.~\cite{4N}, and include the
direct and all 23 exchange terms. Due to the spin-isospin structure,
only 3 topologies contribute to neutron matter: the
three-pion-exchange 4N forces $V^a$ and $V^e$, and the
pion-pion-interaction 4N forces $V^f$. The 4N forces $V^k$ and $V^n$
involving the contact $C_T$ vanish in neutron matter due to their spin
structure. We find a total 4N force contribution of $-174 \pm 10\kev$
per particle at $n_0$. The $V^e$ and $V^f$ energies largely
cancel~\cite{Riska}, and their sum agrees with the very small $\sim
-20 \kev$ per particle at $n_0$ of Ref.~\cite{Fiorilla2011}, which
considered these two parts.

Since diagrams beyond Hartree-Fock involving NN interactions and
N$^2$LO 3N forces (in particular with the larger $c_i$ at N$^3$LO
3N and without RG evolution)
provide non-negligible contributions~\cite{nm}, we include all such
diagrams to second order, as well as particle-particle diagrams to
third order, which is technically possible based on
Ref.~\cite{nucmatt2}. In addition to using NN potentials with
different cutoffs and varying the 3N/4N cutoffs, we include estimates
of the theoretical uncertainties of the $c_i$ constants and in the
convergence of the many-body calculation. The latter is probed by
studying the sensitivity of the energy to the single-particle spectrum
used. We find that the energy changes from second to third order,
employing a free or Hartree-Fock spectrum, by $0.8, 0.4, 1.3 \mev$
($1.4, 0.9, 2.7 \mev$) per particle at $n_0/2$ ($n_0$) for the EGM
450/500, 450/700, EM 500 N$^3$LO potentials, respectively. The
results, which include all these uncertainties, are displayed by the
bands in Fig.~\ref{fig:N3LO}. Understanding
the cutoff dependence and developing improved power counting schemes
remain important open problems in chiral EFT~\cite{powercounting}.
For the neutron matter energy at $n_0$,
our first complete N$^3$LO calculation yields $14.1-21.0 \mev$ per
particle. If we were to omit the results based on the EM 500 N$^3$LO
potential, as it converges slowest at $n_0$, the range would be
$14.1-18.4 \mev$.

\begin{figure}[b]
\begin{center}
\vspace*{-4mm}
\includegraphics[width=0.75\columnwidth,clip=]{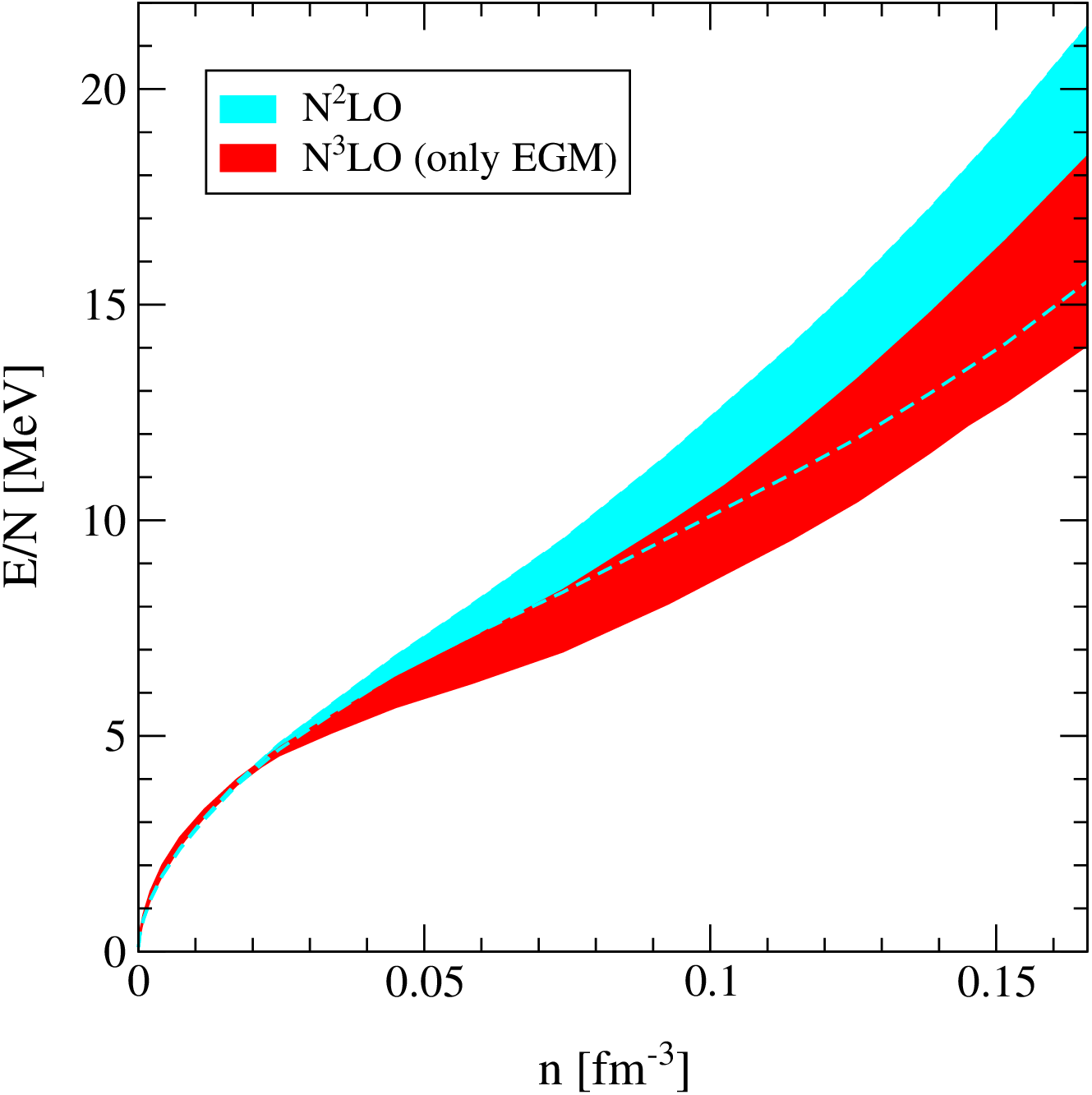}
\vspace*{-4mm}
\end{center}
\caption{(Color online) Neutron matter energy per particle as a
function of density at N$^2$LO (upper/blue band that extends to the
dashed line) and N$^3$LO (lower/red band). The bands are based on
the EGM NN potentials and include uncertainty estimates
as in Fig.~\ref{fig:N3LO}.\label{fig:comparison}}
\end{figure}

As we find relatively large contributions from N$^3$LO 3N forces, it
is important to study the EFT convergence from N$^2$LO to
N$^3$LO. This is shown in Fig.~\ref{fig:comparison} for the EGM
potentials (N$^2$LO is not available for EM), where the N$^3$LO
results are found to overlap with the N$^2$LO band across a $\pm 1.5
\mev$ range around $17 \mev$ at saturation density. As expected from
the net-attractive N$^3$LO 3N contributions in
Fig.~\ref{fig:individual}, the N$^3$LO band yields lower energies. For
the N$^2$LO band, we have estimated the theoretical uncertainties in
the same way, and the neutron matter energy ranges from $15.5-21.4
\mev$ per particle at $n_0$. The theoretical uncertainty is reduced
from N$^2$LO to N$^3$LO to $14.1-18.4 \mev$, but not by a factor $\sim
1/3$ based on the power counting estimate. This reflects the large
$c_i$ 3N contributions at N$^4$LO, and is similar to the convergence
pattern observed in chiral NN potentials~\cite{RMP}.

The neutron matter energy in Fig.~\ref{fig:N3LO} is in very good
agreement with NLO lattice results~\cite{NLOlattice} and Quantum Monte
Carlo simulations~\cite{GC} at very low densities (see also the inset)
and approximately reproduces the scaling $\sim 0.5 \, \frac{3
\kf^2}{10 m}$, which we attribute to effective-range effects
combined with low cutoffs~\cite{dEFT}. At nuclear densities, we
compare our N$^3$LO results with variational calculations based on
phenomenological potentials (APR)~\cite{APR}, which are within the
N$^3$LO band, but do not provide theoretical uncertainties. In
addition, we compare the density dependence with results from
Auxiliary Field Diffusion MC calculations (GCR)~\cite{GCR} based on
nuclear force models adjusted to an energy difference of $32 \mev$
between neutron matter and the empirical saturation point. The density
dependence is similar to the N$^3$LO band, but the GCR results are
higher below $0.05 \fmiq$.

\begin{figure}[t]
\begin{center}
\includegraphics[width=0.85\columnwidth,clip=]{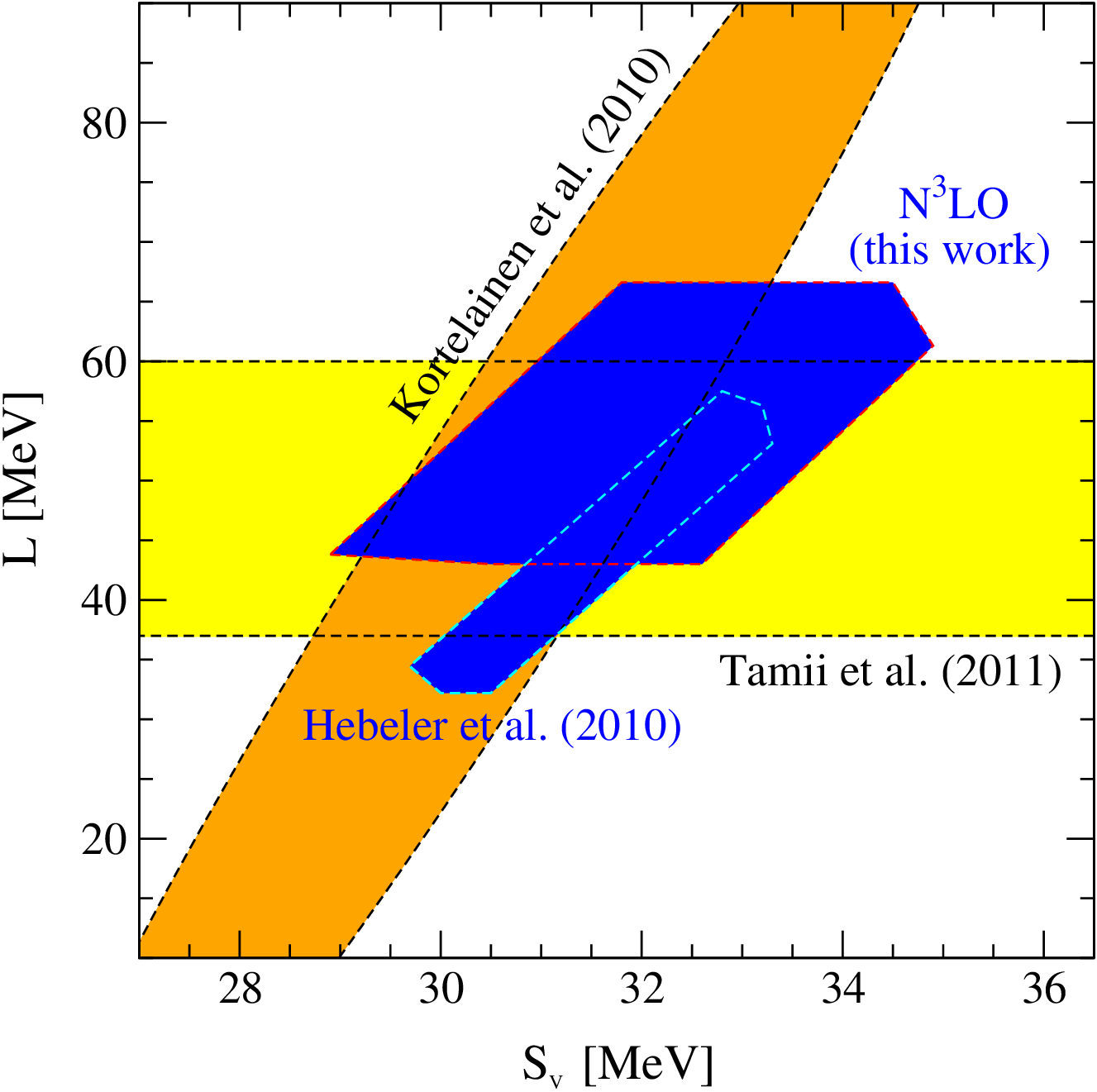}
\vspace*{-4mm}
\end{center}
\caption{(Color online) Range for the symmetry energy $S_v$ and its
density dependence $L$ obtained at N$^3$LO (this work) versus
including 3N forces at N$^2$LO (Hebeler {\it et al.}~\cite{nstar}).
For comparison~\cite{LL}, we show constraints obtained
from energy-density functionals for nuclear masses
(Kortelainen {\it et al.}~\cite{Kortelainen2010}) and from the
$^{208}$Pb dipole polarizability (Tamii {\it et al.}~\cite{Tamii2011}).
\label{fig:SvLband}}
\vspace*{-4mm}
\end{figure}

The N$^3$LO band provides key constraints for the nuclear equation of
state and for astrophysics. Figure~\ref{fig:SvLband} shows, following
Ref.~\cite{LL}, the allowed range for the symmetry energy $S_v$ and
its density dependence $L = 3 n_0 \partial_n S_v(n_0)$ (for details on
the determination of $S_v$ and $L$ see Ref.~\cite{nstar}). Compared to
the results from RG-evolved chiral interactions with 3N forces at
N$^2$LO only~\cite{nstar}, we find the same correlation (with the same
slope), but not as tight due to the additional density dependences at
N$^3$LO. The N$^3$LO ranges for $S_v$ and $L$ are $S_v =
28.9-34.9\mev$ and $L = 43.0-66.6 \mev$. The two neutron-matter bands
in Fig.~\ref{fig:SvLband} are complementary, because the RG evolution
in Hebeler {\it et al.}~\cite{nstar} improves the many-body
convergence, while the band presented in this work is the first
consistent N$^3$LO calculation. The predicted N$^3$LO range, as well
as that of Hebeler {\it et al.}~\cite{nstar}, are in agreement with
constraints obtained from energy-density functionals for nuclear
masses~\cite{Kortelainen2010} and from the $^{208}$Pb dipole
polarizability~\cite{Tamii2011}. In the future, the N$^3$LO band can
be narrowed further by a higher-order many-body calculation with
N$^3$LO 3N forces and by taking into account $\Delta$ excitations
(explicitly or through large $c_i$ contributions at
N$^4$LO~\cite{Krebs2012}). Combined with the heaviest $2 M_\odot$
neutron star~\cite{Demorest} and a general extension to high
densities~\cite{nstar}, our N$^3$LO energy band leads to a radius
range of $R = 9.7 - 13.9$~km for a typical $1.4 M_{\odot}$ neutron
star, in remarkable agreement with Ref.~\cite{nstar}. For an
alternative determination using in-medium chiral perturbation theory
for all densities see Ref.~\cite{Wolfram2}.

We have presented the first complete N$^3$LO calculation of the
neutron matter energy, including NN, 3N and 4N forces, with the first
application of N$^3$LO 3N forces to many-body systems. The significant
contributions from N$^3$LO 3N forces show that their inclusion will also
be very important for nuclear structure and reactions. Our results
provide constraints for the nuclear equation of state and for
neutron-rich matter in astrophysics, and highlight the exciting role
neutron matter and neutron-rich systems play in chiral EFT, where all
many-neutron forces are predicted. The large contributions from
N$^3$LO 3N forces signal the importance of $\Delta$ contributions at
nuclear densities.

We thank E.\ Epelbaum, R.\ J.\ Furnstahl, N.\ Kaiser and H.\ Krebs for
discussions. This work was supported by the Helmholtz Alliance Program
of the Helmholtz Association, contract HA216/EMMI ``Extremes of
Density and Temperature: Cosmic Matter in the Laboratory'', the DFG
through Grant SFB 634, the ERC Grant No.~307986 STRONGINT, and the 
NSF Grant No.~PHY--1002478.

\end{document}